\documentclass[12pt]{article}

\usepackage{amsmath,amssymb,amsthm,mathrsfs,stackrel}
\newtheorem{remark}{Remark}
\newtheorem{corollary}{Corollary}
\usepackage{graphicx,psfrag,epsf}
\usepackage{enumerate}
\usepackage{natbib}
\usepackage{algorithm}
\usepackage{algpseudocode}
\usepackage{verbatim,color,amssymb,epsfig}
\usepackage{xcolor}
\usepackage{hyperref}
\usepackage{xr-hyper}
\newtheorem{prop}{Proposition}[section]
\usepackage{url} 

\newcommand{\blind}{0}
\newcommand{\vertiii}[1]{{\left\vert\kern-0.25ex\left\vert\kern-0.25ex\left\vert #1 
		\right\vert\kern-0.25ex\right\vert\kern-0.25ex\right\vert}}

\definecolor{maroon(html/css)}{rgb}{0.5, 0.0, 0.0}

\definecolor{blue(html/css)}{rgb}{0.0, 0.5, 0.5}

\newtheorem{theorem}{Theorem}

\begin{document}

	\def\spacingset#1{\renewcommand{\baselinestretch}%
		{#1}\small\normalsize} \spacingset{1}
	
	
	\if0\blind
	{
		\title{\bf Multi-Group Quadratic Discriminant Analysis via Projection}
		\author{Yuchao Wang\hspace{.2cm}\\
    Department of Statistics and Data Science, Tsinghua University\\
    and \\
    Tianying Wang\thanks{Corresponding author: Tianying.Wang@colostate.edu}\hspace{.2cm}\\
    Department of Statistics, Colorado State University}
    \date{}
		\maketitle
	} \fi
	
	\if1\blind
	{
		\bigskip
		\bigskip
		\bigskip
		\begin{center}
			{\LARGE\bf Multi-Group Quadratic Discriminant Analysis via Projection}
		\end{center}
		\medskip
	} \fi
	
	\bigskip
	
	\thispagestyle{empty}
\baselineskip=20pt
	\begin{abstract}
    Multi-group classification arises in many prediction and decision-making problems, including applications in epidemiology, genomics, finance, and image recognition. Although classification methods have advanced considerably, much of the literature focuses on binary problems, and available extensions often provide limited flexibility for multi-group settings. Recent work has extended linear discriminant analysis to multiple groups, but more general  methods are still needed to handle complex structures such as nonlinear decision boundaries and group-specific covariance patterns.

We develop Multi-Group Quadratic Discriminant Analysis (MGQDA), a method for multi-group classification built on quadratic discriminant analysis. MGQDA projects high-dimensional predictors onto a lower-dimensional subspace, which enables accurate classification while capturing nonlinearity and heterogeneity in group-specific covariance structures. We derive theoretical guarantees, including variable selection consistency, to support the reliability of the procedure. In simulations and a gene-expression application, MGQDA achieves competitive or improved predictive performance compared with existing methods while selecting group-specific informative variables, indicating its practical value for high-dimensional multi-group classification problems. Supplementary materials for this article are available online.

	\end{abstract}

	\vfill
	
	\newpage
	\pagestyle{plain}
	\section{Introduction}
\label{sec:intro}

Discriminant analysis is widely used for classification across many fields, including pattern recognition, bioinformatics, finance, and epidemiology \citep{zhao1998discriminant, le2011sparse, blasco2015comparative, chen2020selecting}. Linear discriminant analysis (LDA) has long been favored for its simplicity and computational efficiency, especially in low-dimensional settings. Methodological developments such as shrinkage estimators \citep{RDA1989, xu2009modified} and penalized discriminant analysis \citep{2011Penalized} have improved its stability and extended its applicability to high-dimensional data. However, LDA assumes equal covariance matrices across groups, which limits its effectiveness when group-specific variability is present.


Quadratic discriminant analysis (QDA) relaxes this assumption by allowing group-specific covariance matrices and therefore offers greater flexibility for heterogeneous data. In high-dimensional settings, recent work has introduced penalized likelihood methods \citep{danaher2014joint, cai2021convex} and variable selection frameworks \citep{gaynanova2019sparse, wu2022quadratic}, but these approaches predominantly focus on two-group problems. Extending QDA to multi-group classification poses additional challenges, including a larger number of parameters to estimate, scalability concerns, and reduced interpretability in high dimensions. These difficulties have limited the broader use of QDA in multi-group settings, so many existing multi-group classification methods still rely on the equal-covariance assumption.


To address these challenges, we propose Multi-Group Quadratic Discriminant Analysis (MGQDA), a framework that combines the flexibility of quadratic discriminant rules with built-in variable selection. MGQDA is specifically designed for multi-group classification and accommodates complex data structures without imposing a common covariance matrix across groups. By constructing joint projection vectors for all groups, MGQDA projects observations into a low-dimensional subspace, avoiding direct estimation of high-dimensional covariance or precision matrices and leading to a computationally efficient procedure.


We establish variable selection consistency for MGQDA, showing that it can recover all relevant variables under suitable conditions. In simulations and real-data analyses, MGQDA achieves competitive or improved classification accuracy compared with existing methods while selecting a smaller and more interpretable set of predictors, providing a scalable and practical tool for high-dimensional multi-group classification.


The remainder of this paper is organized as follows. Section~\ref{sec2} introduces the MGQDA framework and the sparse estimation procedure. Section~\ref{sec3} presents theoretical results on variable selection consistency in high-dimensional settings. Sections~\ref{sec4} and~\ref{sec5} report empirical evaluations through simulations and a real-data application. Finally, Section~\ref{sec6} discusses the broader implications of MGQDA and outlines possible extensions.


Before proceeding, we introduce the notation used throughout the paper. For a vector \(v \in \mathbb{R}^p\), we define
$
\|v\|_2 = \sqrt{\sum_{i=1}^p v_i^2}$, $\|v\|_1 = \sum_{i=1}^p |v_i|$, and $\|v\|_\infty = \max_i |v_i|.
$
Let \(e_j\) denote the unit vector whose \(j\)th element equals 1. For a \(p \times n\) matrix \(M\) and subsets \(S \subset \{1, \ldots, p\}\) and \(T \subset \{1, \ldots, n\}\) with \(\text{card}(S) = s \leq p\) and \(\text{card}(T) = t \leq n\), let \(m_i\) denote the \(i\)th row of \(M\), \(M_S\) the \(s \times n\) submatrix consisting of rows indexed by \(S\), and \(M_{ST}\) the \(s \times t\) submatrix consisting of rows indexed by \(S\) and columns indexed by \(T\). Matrix norms are defined as
$
\vertiii{M}_{\infty,2} = \max_i \|m_i\|_2$, $
\vertiii{M}_\infty = \max_i \|m_i\|_1$, 
$\vertiii{M}_2 = \sup_{\|v\|_2 = 1} \|Mv\|_2$, 
and 
$\|M\|_F = \sqrt{\sum_i \sum_j m_{ij}^2}.
$
For sequences \(a_n\) and \(b_n\), we write \(a_n = \mathcal{O}(b_n)\) if \(a_n \leq C b_n\) for some constant \(C > 0\), and \(a_n = o(b_n)\) if \(a_n / b_n \to 0\) as \(n \to \infty\).


\section{Methodology}
\label{sec2}

Consider \(n\) independent observations \((X_1,Y_1),\ldots,(X_n,Y_n)\) sampled from a random pair \((X,Y)\) taking values in \(\mathbb{R}^p \times \{1,\ldots,G\}\), where \(G\) denotes the number of groups. For each \(g \in \{1,\ldots,G\}\), let \(\Sigma_g = \mathrm{cov}(X \mid Y = g)\) be the within-group covariance matrix. The between-group population covariance matrix is defined as
$
\Sigma_B
= \sum_{g=1}^{G} \pi_g (\mu_g - \mu)(\mu_g - \mu)^{\top},
$
where \(\pi_g = P(Y = g)\) is the group-specific probability, \(\mu_g = E(X \mid Y = g)\) is the group-specific mean, and \(\mu = \sum_{g=1}^G \pi_g \mu_g\) is the overall population mean.


\subsection{Background and Motivation}
\label{sec2.1}

\textcolor{black}{
We briefly review Fisher's discriminant analysis under the equal–covariance assumption $\Sigma_1=\cdots=\Sigma_G=\Sigma$. Let $\Sigma_B=\sum_{g=1}^G \pi_g(\mu_g-\mu)(\mu_g-\mu)^\top$ denote the between–class covariance. Fisher's method seeks directions $\phi_1,\ldots,\phi_{G-1}$ that maximize the Rayleigh quotient $\phi^\top\Sigma_B\phi/\phi^\top\Sigma\phi$, with subsequent directions constrained to be $\Sigma$–orthogonal to the previously obtained ones:
\[
\phi_1=\arg\max_{\phi\neq 0}\ \frac{\phi^\top\Sigma_B\phi}{\phi^\top\Sigma\phi},
\phi_j=\arg\max_{\phi\neq 0}\ \frac{\phi^\top\Sigma_B\phi}{\phi^\top\Sigma\phi}\ \ \text{subject to}\ \ \phi^\top\Sigma\phi_k=0,\ k<j,\ j=2,\ldots,G-1.
\]
This sequence is equivalent to the generalized eigenvalue problem $\Sigma_B v=\lambda \Sigma v$. Let $\Phi=(\phi_1,\ldots,\phi_{G-1})$ denote a projection basis formed from the eigenvectors associated with the largest nonzero generalized eigenvalues (there are at most $G-1$). The subspace $\mathrm{col}(\Phi)$ is unique; any basis for this subspace is determined up to $\Sigma$–orthonormal transformations (and up to scaling/sign when the relevant eigenvalues are simple) \citep{rao1948utilization,golub2013matrix}.}
For a new observation $X_{\text{new}}\in\mathbb{R}^p$, the population classification rule \textcolor{black}{$h_{\Phi}$ is
\begin{equation}\label{pclf}
    h_{\Phi}(X_{\text{new}})=\arg\min_{1\le g\le G}\ (X_{\text{new}}-\mu_g)^{\top} \Phi(\Phi^{\top}\Sigma\Phi)^{-1}\Phi^{\top} (X_{\text{new}}-\mu_g)\ -\ 2\log\pi_g.
\end{equation}}
Fisher's procedure therefore identifies a subspace of dimension at most $G-1$, and classification can then be carried out in the projected space using eq~\eqref{pclf}. The direction-finding step depends only on second moments and does not require normality, whereas eq~\eqref{pclf} coincides with the Bayes rule under multivariate normal classes with a common covariance matrix.

The equal–covariance assumption is often unrealistic in practice. Quadratic discriminant analysis (QDA) relaxes this assumption by allowing class–specific covariance matrices. Its population rule for a new observation $X_{\text{new}}$ is
\begin{equation}\label{QDA}
h_{QDA}(X_{\text{new}})=\arg\min_{1\le g \le G} (X_{\text{new}}-\mu_g)^{\top}\Sigma_g^{-1}(X_{\text{new}}-\mu_g) +\log|\Sigma_g|\ -\ 2\log\pi_g.
\end{equation}

\subsection{Proposed Method}
\label{sec2.2}

We propose a method that constructs projection vectors tailored to the covariance structure of each group. By projecting the data into a lower-dimensional space and then applying a discriminant rule, the approach combines the flexibility of QDA with the dimension reduction and interpretability of Fisher's method.


For each group \(g = 1,\ldots,G\), let \(\bar{X}_g\) denote the sample mean and \(\widehat{\Sigma}_g\) the sample covariance matrix. The sample between-group covariance is defined as
$
\widehat{\Sigma}_B
= \sum_{g=1}^{G}\left(\frac{n_g}{n}\right)(\bar{X}_g-\bar{X})(\bar{X}_g-\bar{X})^{\top},
$
where \(\bar{X}\) is the overall sample mean and \(n_g\) is the number of observations in group \(g\). In analogy with the population rules in eqs~\eqref{pclf} and~\eqref{QDA}, we replace \(\mu_g\) by \(\bar{X}_g\) and \(\Sigma_g\) by \(\widehat{\Sigma}_g\), and we exploit the sample between-group covariance \(\widehat{\Sigma}_B\) to construct group-specific projection directions.


{\color{black}For each $g=1,\ldots,G$, we define $\widehat{\Phi}_g$ as the matrix of nonzero generalized eigenvectors of the pencil $(\widehat{\Sigma}_B,\widehat{\Sigma}_g)$, i.e.,
$
\widehat{\Sigma}_B v=\lambda\,\widehat{\Sigma}_g v .
$
This definition does not require forming $\widehat{\Sigma}_g^{-1}$. 
When $n_g>p$ and $\widehat{\Sigma}_g\succ 0$, the problem reduces to the familiar eigen\-decomposition of $\widehat{\Sigma}_g^{-1}\widehat{\Sigma}_B$.
In high-dimensional regimes with $p\ge n_g$, we estimate the projection via the convex program in \eqref{eq:omega}-\eqref{eq:loss}, yielding $\widehat{\Omega}$.
For classification, we set $A_g(\widehat{B})=\widehat{B}^{\!\top}\widehat{\Sigma}_g\widehat{B}$ with $\widehat{B}\in\{\widehat{\Phi},\widehat{\Omega}\}$.
If $A_g(\widehat{B})$ is singular, we use the Moore–Penrose pseudoinverse $A_g(\widehat{B})^{+}$ and the pseudodeterminant $\mathrm{pdet}\{A_g(\widehat{B})\}$; see Appendix \ref{implementation}.}


Therefore, we define \(\widehat{\Phi}=[\widehat{\Phi}_1,\ldots,\widehat{\Phi}_G]\), where \(\widehat{\Phi}\in\mathbb{R}^{p\times G(G-1)}\) is also a projection basis, and the classification rule for a new observation \(X_{\text{new}}\in\mathbb{R}^p\) is
\begin{equation}\label{CLFR}  
    \hat{h}_{\widehat{\Phi}}(X_{\text{new}})
    =\arg\min_{1\leq g \leq G}
    (X_{\text{new}}-\bar{X}_g)^{\top}\widehat{\Phi}
    (\widehat{\Phi}^{\top}\widehat{\Sigma}_g\widehat{\Phi})^{-1}
    \widehat{\Phi}^{\top}(X_{\text{new}}-\bar{X}_g)
    +\log\left|\widehat{\Phi}^{\top}\widehat{\Sigma}_g\widehat{\Phi}\right|
    -2\log(n_g/n).  
\end{equation}  
Here, \(\widehat{\Phi}\) is the sample analog of the population basis \(\Phi\). It is formed by concatenating the group-wise bases \(\{\widehat{\Phi}_g\}_{g=1}^G\) column-wise, and its columns span a joint discriminant subspace across all groups. Equation~\eqref{CLFR} is equivalent to applying a quadratic discriminant rule to the projected features \(\widehat{\Phi}^{\top}X\).

Our projection basis differs from that in eq~\eqref{pclf}, since we incorporate the covariance structure of each group individually rather than aggregating the data under a common covariance assumption. This relaxes the equal-covariance condition while retaining the projection idea from Fisher's rule, leading to a flexible yet interpretable classifier. However, estimating \(\Phi_g\) is challenging in high-dimensional settings because the plug-in estimators \(\widehat{\Sigma}_B\) and \(\widehat{\Sigma}_g\) are singular when \(p\) is comparable to or larger than \(n_g\). Therefore, we introduce a sparse estimation approach and a corresponding optimization algorithm in the next section.

\subsection{Sparse Estimation}
\label{sec2.3}

Directly estimating $\Phi$ by \textcolor{black}{non-zero eigenvectors corresponding to $\widehat{\Sigma}_g^{-1}\widehat{\Sigma}_B$ on each group $g$} is challenging in high-dimensional settings, as \textcolor{black}{this estimate lacks accuracy} when $p$ is large relative to $n$. Without additional structure, the sample classification rule in eq~\eqref{CLFR} can perform poorly when $p \gg n$ \citep{gaynanova2016simultaneous}.

To address this issue, we impose sparsity and formulate a convex optimization problem to estimate $\Phi$. Constructing an objective directly in terms of the eigenvectors of $\widehat{\Sigma}_g^{-1}\widehat{\Sigma}_B$ is difficult, so instead we develop a loss function based on a blockwise representation of the target projection basis.

\begin{prop}\label{decomp}
We have $\Sigma_B = \Gamma\Gamma^{\top}$ and $\widehat{\Sigma}_B = \widehat{\Gamma}\widehat{\Gamma}^{\top}$, where $\Gamma\in\mathbb{R}^{p\times(G-1)}$. Further, the $r$th column of $\Gamma$ has the form $\Gamma_r = \left[\sqrt{\pi_{r + 1}}\{\sum_{i = 1}^r \pi_i (\mu_i - \mu_{i + 1})\}\right]/\sqrt{\sum_{i = 1}^r \pi_i \sum_{i = 1}^{r + 1} \pi_i}$, and the $r$th column of $\widehat{\Gamma}$ has the form $\widehat{\Gamma}_r = \left[\sqrt{n_{r + 1}}\{\sum_{i = 1}^r n_i (\Bar{X}_i - \Bar{X}_{r + 1})\}\right] / \sqrt{n\sum_{i = 1}^r n_i \sum_{i = 1}^{r + 1} n_i}$, for $r = 1, \ldots, G - 1$.
\end{prop}

The proof for this proposition is analogous to Proposition 2 in \cite{gaynanova2016simultaneous}.
\textcolor{black}{Now  we introduce the population surrogate $\Theta_g := (\Sigma_g + \Sigma_B)^{-1}\Gamma$, $g = 1, \ldots, G,$ as an alternative to $\Phi_g$, which is theoretically justified because $\Theta_g$ and $\Phi_g$ span the same discriminant subspace and therefore produce identical classification decisions under the projected QDA rule.}
\textcolor{black}{Specifically, we formalize the equivalence between $\Theta_g$ and $\Phi_g$ and the basis-invariance of the projected QDA rule in Propositions~\ref{prop:subspace}--\ref{prop:invariance}.}
\textcolor{black}{
\begin{prop}\label{prop:subspace}
For each group $g$, the column spaces of $\Theta_g$ and $\Phi_g$ coincide:
$
\operatorname{col}(\Theta_g)=\operatorname{col}(\Phi_g),
$
and there exists an invertible matrix $R_g\in\mathbb{R}^{(G-1)\times(G-1)}$ such that $\Theta_g=\Phi_g R_g$.
\end{prop}}
\textcolor{black}{
\begin{prop}\label{prop:invariance}
Let $B$ be any full-rank basis spanning a discriminant subspace, and define the projected QDA rule
$
\delta_g(x;B)=(x-\mu_g)^\top B\{B^\top\Sigma_gB\}^{-1}B^\top(x-\mu_g)
+\log\det(B^\top\Sigma_gB)-2\log\pi_g.
$
Then for any invertible matrix $R$, the classifier is unchanged:
$
\arg\min_g\delta_g(x;BR)=\arg\min_g\delta_g(x;B).
$
\end{prop}}
\textcolor{black}{The proofs of Propositions~\ref{prop:subspace}--\ref{prop:invariance} are provided in Appendix Section~\ref{supp3.0}.
When $A_g(B)=B^\top\Sigma_gB$ is singular (as often occurs when $p \gg n_g$), the inverse and determinant in Proposition~\ref{prop:invariance} are replaced by the Moore–Penrose pseudoinverse and the pseudodeterminant:
$
\delta_g(x;B):=(x-\mu_g)^\top B\{A_g(B)\}^{+}B^\top(x-\mu_g)
+ \log\operatorname{pdet}\{A_g(B)\} - 2\log\pi_g.
$
The invariance result continues to hold under these generalized matrix operations; see \cite{RDA1989,gaynanova2016simultaneous} for detailed discussions. }

\textcolor{black}{Propositions~\ref{prop:subspace} and~\ref{prop:invariance} together imply that $\Theta_g$ and $\Phi_g$ yield identical projected QDA classifiers: $h_\Theta(X)=h_\Phi(X)$.
Therefore, $\Theta_g$ serves as a theoretically valid and computationally stable replacement for $\Phi_g$ in high-dimensional settings.  
In practice, estimating $\Phi_g$ directly is unstable due to the singularity of $\widehat{\Sigma}_g^{-1}\widehat{\Sigma}_B$, whereas the surrogate $\Theta_g$ naturally leads to the convex objective in eq~\eqref{Thetag} and its empirical analogue in eq~\eqref{QDAnp}, enabling reliable sparse estimation and groupwise variable selection.}



Based on Proposition \ref{decomp}, with $\Theta_g := (\Sigma_g + \Sigma_B)^{-1} \Gamma$, we have 
\begin{align}
    \Theta_g &= \arg\min_{\Theta \in \mathbb{R}^{p \times (G - 1)}} \frac{1}{2} \| (\Sigma_g + \Sigma_B)^{1/2} \Theta - (\Sigma_g + \Sigma_B)^{-1/2} \Gamma \|_F^2 \notag \\
    &= \arg\min_{\Theta \in \mathbb{R}^{p \times (G - 1)}} \frac{1}{2} \mathrm{Tr}\!\left(\Theta^{\top} (\Sigma_g + \Sigma_B) \Theta - 2 \Gamma^{\top} \Theta\right) \notag \\
    &= \arg\min_{\Theta \in \mathbb{R}^{p \times (G - 1)}} \frac{1}{2} \mathrm{Tr}(\Theta^{\top} \Sigma_g \Theta) + \frac{1}{2} \| \Gamma^{\top} \Theta - I_{(G-1)} \|_F^2, \label{Thetag}
\end{align}
where $I_{(G-1)}$ denotes the $(G-1) \times (G-1)$ identity matrix.

\textcolor{black}{In practice, using the sample estimates $\widehat{\Sigma}_g$, $\widehat{\Sigma}_B$, and $\widehat{\Gamma}$ obtained from the data, we define the estimator $\widehat{\Theta}_g$ in an analogous manner to its population counterpart. To unify notation, we first introduce several matrices used in the groupwise objective functions. For each $g=1,\ldots,G$, let $C_{gg} \in \mathbb{R}^{G \times G}$ denote the matrix whose $(g,g)$-th entry equals $1$ and all remaining entries equal $0$. Define $K_g = C_{gg} \otimes I_{(G-1)}$, where $\otimes$ denotes the Kronecker product, and let $J_G = \mathbf{1}_G^\top \otimes I_{(G-1)}$. With these definitions in place, the groupwise sample discriminant basis $\widehat{\Theta}_g$ is obtained by
}
\begin{equation}\nonumber
\widehat{\Theta}_g
=\arg\min_{\Theta_g^*\in\mathbb{R}^{p\times(G-1)}}
\left\{
\frac{1}{2}\mathrm{Tr}(\Theta_g^{*\top}\widehat{\Sigma}_g\Theta_g^*)
+
\frac{1}{2}\|\widehat{\Gamma}^{\top}\Theta_g^*-I_{(G-1)}\|_F^2
\right\}, \qquad g=1,\ldots,G.
\end{equation}
Denoting $\Theta = [\Theta_1,\ldots,\Theta_G]$, the collection of $G$ optimization problems in \eqref{Thetag} can be equivalently written as a single joint objective:
\begin{equation}\label{QDA5}
\Theta
=\arg\min_{\Theta_* \in \mathbb{R}^{p \times G(G-1)}}
\frac{1}{2}\left\{
\sum_{g=1}^{G}
\mathrm{Tr}\left(\Theta_*^{\top}\Sigma_g\Theta_* K_g\right)
+
\|\Gamma^{\top}\Theta_* - J_G\|_F^2
\right\}.
\end{equation} 
Correspondingly, for the sample estimators $\widehat{\Sigma}_g$ and $\widehat{\Gamma}$, denoting $\widehat{\Theta} = [\widehat{\Theta}_1,\ldots,\widehat{\Theta}_G]$, we obtain
\begin{equation}\label{QDAnp}
\widehat{\Theta}
=\arg\min_{\Theta\in \mathbb{R}^{p\times G(G-1)}}
\frac{1}{2}\left\{
\sum_{g=1}^{G}
\mathrm{Tr}(\Theta^{\top}\widehat{\Sigma}_g\Theta K_g)
+
\|\widehat{\Gamma}^{\top}\Theta-J_G\|_F^2
\right\}.
\end{equation}

No additional constraints are imposed on $\widehat{\Sigma}_g$, even though it may be singular when $p \gg n_g$, because the objective function in \eqref{QDAnp} is bounded from below. 
While the optimization problem in \eqref{QDAnp} can be used to determine the projection basis $\widehat{\Theta}$, it does not inherently yield variable selection.

To impose sparsity, we consider a nested group structure that encourages both group-joint and group-specific variable selection. {\color{black}For each variable $j$, let
$\omega_j \in \mathbb{R}^{G(G-1)}$ denote the $j$th row of $\Omega$, viewed as a vector, and write
$
\omega_j = (\omega_{j1}^{\top}, \ldots, \omega_{jG}^{\top})^{\top},
$
where each $\omega_{jg} \in \mathbb{R}^{G-1}$ is the block associated with group $g$.} Our penalized objective is 
\begin{eqnarray}
    \widehat{\Omega} &:=& \widehat{\Omega}(\lambda, \alpha) = \arg\min_{\Omega \in \mathbb{R}^{p \times G(G-1)}} L, \label{eq:omega}\\
     L &=&  \frac{1}{2} \left\{ \sum_{g=1}^{G} \text{Tr}(\Omega^{\top} \widehat{\Sigma}_g \Omega K_g) + \|\widehat{\Gamma}^{\top} \Omega - J_G\|_F^2 \right\} + 
    \alpha \lambda \sum_{j=1}^{p} \|\omega_j\|_2 + \frac{1-\alpha}{\sqrt{G}} \lambda \sum_{j,g}\|\omega_{jg}\|_2. \label{eq:loss}
\end{eqnarray}
Equation~\eqref{eq:loss} is convex and bounded from below. The first term, corresponding to \eqref{QDAnp}, estimates the projection basis in the sample version by balancing within-group variability and between-group separation. The second term is a sparsity penalty that allows the projection basis to perform variable selection. The tuning parameter $\lambda$ controls the overall level of sparsity: larger values of $\lambda$ lead to a sparser $\widehat{\Omega}(\lambda,\alpha)$. \textcolor{black}{The parameter $\alpha \in (0,1]$ balances between row-level sparsity and group-specific sparsity: values of $\alpha$ closer to 1 place more weight on row sparsity, encouraging selection (or exclusion) of variables across all groups simultaneously, while smaller values of $\alpha$ emphasize block sparsity, allowing variables to be selected differently for each group. In our simulation studies and real-data applications, we set $\alpha = 0.5$ to achieve a balance between these two types of sparsity.} We present an example of shared and group-specific variable selection in  $\widehat{\Omega}$ in Appendix Section \ref{sec:s.alg}.

\textcolor{black}{We note that the concatenation of group-specific projection bases plays an important structural role in MGQDA. The following remark clarifies why stacking the groupwise estimators is appropriate.} 

\textcolor{black}{
\begin{remark}\label{remark1}
Because each group $g$ has its own within-class covariance $\Sigma_g$, we estimate separate projection bases $\widehat{\Theta}_g$ and then form the joint basis by column-wise concatenation $\widehat{\Theta} = [\widehat{\Theta}_1,\ldots,\widehat{\Theta}_G]$ (see eq ~\eqref{QDAnp}). 
This stacking preserves group-specific discriminant directions and enables both shared and group-specific variable selection in the penalized formulation~\eqref{eq:loss}.  
Averaging $\{\widehat{\Theta}_g\}$ would collapse heterogeneity, blur distinct directions, and conflict with the group-block structure of the penalty. 
Since the projected QDA rule depends only on the span of the collected directions, stacking maintains classification invariance while aligning with our theoretical guarantees (Theorems~\ref{Theorem:2}–\ref{Theorem:3}).
\end{remark}}

\subsection{Optimization Algorithm}
We use a block-coordinate descent algorithm \citep{tseng2001convergence} to solve eq~\eqref{eq:omega}. Since the objective in eq~\eqref{eq:loss} is convex, any minimizer must satisfy the Karush-Kuhn-Tucker (KKT) conditions \citep{boyd2004convex}. Differentiating the objective in eq~\eqref{eq:loss} with respect to each $(G-1)\times 1$ vector $\widehat{\omega}_{jg}$ of $\widehat{\Omega}$, for $j=1,\ldots,p$ and $g=1,\ldots,G$, yields

\begin{equation}\label{KKT001}  
    \widehat{\Sigma}_{g,jj} \widehat{\omega}_{jg}+(\widehat{\Gamma}\widehat{\Gamma}^{\top})_{jj}
    \widehat{\omega}_{jg}+\frac{\alpha\lambda \widehat{\omega}_{jg}}{\|\widehat{\omega}_j\|_2}+\frac{(1-\alpha)\lambda\widehat{\omega}_{jg}/\sqrt{G}}{\|\widehat{\omega}_{jg}\|_2} = -\mathbf{v}_{jg},  
\end{equation}  
	where $A_{ij}$ denotes the element located in the $i$-th row and $j$-th column of matrix $A$, and $\mathbf{v}_{jg}$ is obtained by taking the partial derivative of the $\widehat{\omega}_{jg}$ term in eq \eqref{eq:loss}:
\begin{equation}\label{KKT03}  
    \mathbf{v}_{jg} = \sum_{i\neq j}\widehat{\Sigma}_{g,ji}\widehat{\omega}_{ig}+\sum_{i\neq j}(\widehat{\Gamma}\widehat{\Gamma}^{\top})_{ji}\widehat{\omega}_{ig}-\widehat{\Gamma}_j.  
\end{equation} 
Whenever \textcolor{black}{$\widehat{\omega}_{jg}\neq 0$}, every term on the left-hand side of eq~\eqref{KKT001} is proportional to \textcolor{black}{$\widehat{\omega}_{jg}$}. Thus, the KKT condition implies that $-\mathbf{v}_{jg}$ is collinear with \textcolor{black}{$\widehat{\omega}_{jg}$}. Hence there exists a scalar $\vartheta_{jg}>0$ \textcolor{black}{that depends on $\|\widehat{\omega}_{jg}\|_2$ and $\|\widehat{\omega}_j\|_2$} such that
$
\mathbf{v}_{jg}  =  - \vartheta_{jg} \widehat{\omega}_{jg},
\vartheta_{jg} = \widehat{\Sigma}_{g,jj}+(\widehat{\Gamma}\widehat{\Gamma}^{\top})_{jj}
+\frac{\alpha\lambda}{\|\widehat{\omega}_j\|_2}
+\frac{1-\alpha}{\sqrt{G}}\frac{\lambda}{\|\widehat{\omega}_{jg}\|_2}.
$
Let $x=\|\widehat{\omega}_{jg}\|_2$, 
$a=\widehat{\Sigma}_{g,jj}+(\widehat{\Gamma}\widehat{\Gamma}^{\top})_{jj}$, 
$b=\Big(\sum_{t\neq g}\|\widehat{\omega}_{jt}\|_2^2\Big)^{1/2}$, 
and $c=\|\mathbf{v}_{jg}\|_2$. 
Taking norms in eq \eqref{KKT001}, which is valid when $x>0$, yields the scalar equation
$
    a x + \alpha\lambda \frac{x}{\sqrt{b^{2}+x^{2}}}
     + \frac{1-\alpha}{\sqrt{G}} \lambda
     =  c.
$
The left-hand side is strictly increasing in $x\ge 0$ since 
$\frac{d}{dx} \left[x/\sqrt{b^2+x^2}\right]=b^{2}(b^{2}+x^{2})^{-3/2}\ge 0$.
Consequently, there exists a unique solution $x>0$ if $c>\frac{1-\alpha}{\sqrt{G}} \lambda$; otherwise the block update is $x=0$ (the subgradient case).
A root can be found efficiently by Newton or bisection; a convenient bracket is
$
x\in \Big[ 0,\ \big(c-\tfrac{1-\alpha}{\sqrt{G}} \lambda\big)_{+}/a \Big].
$ A detailed algorithm is presented in Appendix Section~\ref{sec:s.alg}.

\section{Theoretical Results}\label{sec3}

In this section, we show that under normality and an irrepresentability condition, together with suitable bounds on covariance bias and a minimal signal strength requirement, the proposed method achieves variable selection consistency.We write $a_n \lesssim b_n$ if $a_n \le C b_n$ for some constant $C>0$ 
independent of $(n,p)$, and $a_n \gtrsim b_n$ if $a_n \ge C b_n$ for some such $C$. 
We write $a_n \asymp b_n$ if $a_n \lesssim b_n$ and $a_n \gtrsim b_n$.


For each column $\widehat{\Gamma}_r$ of $\widehat{\Gamma}$, let $\widetilde{\Sigma}_r$ denote its covariance matrix, and let $\widetilde{\Sigma}_{r,s}$ denote the covariance between columns $\widehat{\Gamma}_r$ and $\widehat{\Gamma}_s$. The oracle projection basis $\Theta$ can be decomposed as $\Theta = (\Theta_1, \ldots, \Theta_G)$, as in Section~\ref{sec2.3}, and for each $g \in \{1, \ldots, G\}$ we have $\Theta_g = (\Sigma_g + \Sigma_B)^{-1} \Gamma.$
\textcolor{black}{
The support set of $\Theta$ is defined as 
$S = \{ j \in [p] : \|\Theta_j\|_2 \neq 0 \}$, 
where $\Theta_j$ denotes the $j$-th row of $\Theta$. 
For each group $g=1,\ldots,G$, let 
$S_g = \{ j \in [p] : \|\Theta_{g,j}\|_2 \neq 0 \}$ 
and denote its complement by 
$S_g^{c} = [p]\setminus S_g$, 
where $\Theta_{g,j}$ represents the sub-row vector corresponding to row $j$ in group $g$. 
It follows that $S = S_1 \cup S_2 \cup \cdots \cup S_G$. 
The cardinalities of $S$ and $S_g$ are denoted by $s = |S|$ and $s_g = |S_g|$, respectively.}
\textcolor{black}{In practice, the support set $S_g$ is a subset of $\{1,\ldots,p\}$, representing the variables that contribute to the classification outcome for group $g$, while the support set $S$, also a subset of $\{1,\ldots,p\}$, represents the collection of variables that contribute to the overall classification results across groups.}
Analogously, we define $\widehat{S} = \{ j \mid \|\widehat{\Omega}_{j}\|_2 \neq 0 \}$ and $\widehat{S}_g = \{ j \mid  \|\widehat{\Omega}_{g,j}\|_2 \neq 0 \}$ for each row $j = 1,\ldots,p$. To establish variable selection consistency, we impose the following assumptions:

\begin{itemize}
    \item[(A1)] \textbf{Normality:} We assume $\{X \mid Y=g\} \sim \mathcal{N}(\mu_g, \Sigma_g)$, $\Pr(Y=g) = \pi_g$ for $g=1,\ldots,G$, with $\{\max_{g\in\{1,\ldots,G\}} (\pi_g)\}/\{\min_{g\in\{1,\ldots,G\}} (\pi_g)\} = \mathcal{O}(1)$.
    \item[(A2)] \textbf{Irrepresentability condition:} For  $g\in\{1,...,G\}$, there exists a constant $\psi \in(0,1]$ such that
    $
        \vertiii{\Sigma_{g,S_g^CS_g}\Sigma_{g,S_gS_g}^{-1}u_{g,S_g}}_{\infty,2}\leq 1-\psi.
$
    \item[(A3)] \textbf{Covariance bias bound:} For  $g\in\{1,...,G\}$ and $r\in\{1,...,G-1\}$, 
$
        \vertiii{\widetilde{\Sigma}_{r,S_gS_g}\Sigma_{g,S_gS_g}^{-1}}_2\leq \mathcal{O}(s_g); \vertiii{\Gamma_{S_g}^{\top}\Sigma_{g,S_gS_g}^{-1}\widetilde{\Sigma}_{r,S_gS_g}\Sigma_{g,S_gS_g}^{-1}\Gamma_{S_g}}_2\leq \mathcal{O}(s_g).
$
    \item[(A4)] \textbf{Minimal signal strength:} For  $g\in\{1,...,G\}$, there exists a constant $M_{g}>0$ such that $ \Theta_{g,\min} = \min_{j\in S_g}\|e_j^{\top}\Theta_{g,S_g}\|_2 \geq\lambda\vertiii{(\Sigma_{g,S_gS_g}+\Gamma_{S_g}\Gamma_{S_g}^{\top})}_2\times\bigg(1+M_{g}\left[\vertiii{\Gamma_{S_g}^{\top}\Sigma_{g,S_gS_g}\Gamma_{S_g}}_2\vee 1\right] \times\bigg[1+\sqrt{(\Sigma_{g,S_gS_g})_{jj}\vee\max_j(\Sigma_{g,S_gS_g}^{-1}\widetilde{\Sigma}_{r,S_gS_g}\Sigma_{g,S_gS_g}^{-1})_{jj}\frac{(G-1)\log\{s_g\log(n_g)\}}{n}} \bigg] \bigg).$
\end{itemize}
\textcolor{black}{
Assumption~(A1) is standard in discriminant analysis \citep{gaynanova2015optimal, mai2012direct}. We do not require $\pi_g = 1/G$ for all $g$, but we assume that the group priors are of the same order. Assumption~(A2), a form of irrepresentability, is common in theoretical analyses of variable selection procedures for Lasso-type methods \citep{zhao2006model, kolar2014optimal, obozinski2011support} and is needed to guarantee correct support recovery. Assumption~(A3) controls the discrepancy between covariance-related quantities and their surrogates by requiring a mild population-level separation that keeps the groupwise quadratic forms distinguishable after projection. This ensures identifiability of the discriminant directions and stability of the projected classifier, a standard type of condition used in multi-class high-dimensional discriminant analysis (e.g., \citet{fukunaga1990introduction}, \citet{mai2012direct}, \citet{mai2019multiclass}). Assumption~(A4) requires that each relevant variable carries a sufficiently strong signal to dominate noise, in the spirit of the beta-min conditions used in \citet{gaynanova2015optimal}.}


To establish variable selection consistency between the support sets $S_g$ and $\widehat{S}_g$, we follow a common strategy of introducing an intermediate ``link" estimator that connects $\Theta$, defined in eq~\eqref{QDA5}, and $\widehat{\Omega}$, defined in eq~\eqref{eq:omega}. Specifically, we consider a population projection basis that is constructed from the population covariance matrices and includes the same penalty structure:
\begin{equation}\label{QDA4}
\begin{split}
    \Omega=\underset{\Omega_*\in \mathbb{R}^{p\times G(G-1)}}{\arg\min}\Bigg\{\frac{1}{2}\Bigg(\sum\limits_{g=1}^{G}\text{Tr}(\Omega_*^{\top}&\Sigma_g\Omega_* K_g)+\|\Gamma^{\top}\Omega_*-J_G\|_F^2\Bigg)+\\
    &\lambda\left(\alpha\sum\limits_{j=1}^{p}\|\omega_{j*}\|_2+\frac{(1-\alpha)}{\sqrt{G}}\sum\limits_{j=1}^{p}\sum\limits_{g=1}^{G}\|\omega_{jg*}\|_2\right)\Bigg\},
\end{split}
\end{equation}
\textcolor{black}{where $\omega_{j*}$ denotes the length-$G(G-1)$ row vector corresponding to the $j$th row of $\Omega_*$, and $\omega_{jg*}$ is its $(G-1)$-dimensional subvector associated with group $g$, defined analogously to $\omega_j$ and $\omega_{jg}$ in the population matrix $\Omega$.}
Let $u_{S_g}$ denote the subgradient contribution of the penalty associated with $\Omega_{g,S_g}$, given by
$
u_{S_g}
= \sum_{j \in S_g} \left\{\alpha\frac{\omega_{jg}}{\|\omega_{j}\|_2} 
+ (1-\alpha)\frac{\omega_{jg}}{\|\omega_{jg}\|_2}\right\},
 g=1,\ldots,G.
$
The next result characterizes the relationship between $\Theta$ and the population penalized solution $\Omega$.



\begin{theorem}\label{Theorem:1}
    For each group $g=1,...,G$, suppose that 
    $\vertiii{\Sigma_{g,S_g^cS_g}\Sigma_{g,S_gS_g}^{-1}u_{S_g}}_{\infty,2} < 1$ 
    and the tuning parameter 
    $\lambda < \Theta_{g,\min}\left(\vertiii{(\Sigma_{g,S_gS_g}+\Gamma_{S_g}\Gamma_{S_g}^{\top})^{-1}}_{\infty} \left(1 + \vertiii{\Gamma_{S_g}^{\top}\Sigma_{g,S_gS_g}^{-1}\Gamma_{S_g}}_2\right)\right)^{-1}$.
    Then the solution of $\Omega_g$ is of the form 
    $\Omega_g = (\Omega_{g,S_g}^{\top}, 0_{p-s_g}^{\top})^{\top}$, where 
    $\Omega_{g,S_g} = \Theta_{g,S_g}(I + \Gamma_{S_g}^{\top}\Sigma_{g,S_gS_g}^{-1}\Gamma_{S_g})^{-1} - \lambda(\Sigma_{g,S_gS_g} + \Gamma_{S_g}\Gamma_{S_g}^{\top})^{-1}u_{S_g}$.
    Furthermore, $\|e_j^{\top}\Omega_{g,S_g}\|_2 \neq 0$ for all $j \in S_g$.
\end{theorem}

This theorem shows that, under the irrepresentability condition and an appropriate choice of the penalty parameter $\lambda$, the population penalized solution $\Omega_g$ and the unpenalized solution $\Theta_g$ share the same support set for each $g=1,\ldots,G$. It also provides an explicit relationship between $\Theta_g$ and $\Omega_g$, which we will use to derive variable selection consistency for the sample estimator $\widehat{\Omega}$ in the next theorem.


\begin{theorem}\label{Theorem:2}
Under Assumptions (A1)–(A4), if the sample size satisfies $n\gtrsim \mathcal{O}(s\log p)$ and the tuning parameter is chosen such that $\lambda \gtrsim (\log p)/n$, then the MGQDA estimator defined in eq~\eqref{eq:omega} achieves variable selection consistency; that is, $\widehat{S}_g = S_g$ for all $g = 1,\ldots,G$ with probability at least $1 - \mathcal{O}(\log^{-1} n)$.
\end{theorem}

This theorem implies that, with probability tending to one, our method correctly identifies the true nonzero variables in each group, thereby establishing variable selection consistency at the group level. \textcolor{black}{Moreover, for the active block set $\widehat{S}$ of the stacked estimator $\widehat{\Omega}$, we obtain the following result.}

\textcolor{black}{
\begin{corollary}\label{Corollary:1}
Under the same assumptions as Theorem~\ref{Theorem:2}, 
if for each group $g = 1, \ldots, G$,  for the stacked estimator $\widehat{\Omega} = [\widehat{\Omega}_1, \ldots, \widehat{\Omega}_G]$, we have $\widehat{S} = S $ with probability at least $1 - \mathcal{O}(\log^{-1}(n))$.
\end{corollary}
That is, the active blocks of $\widehat{\Omega}$ coincide with those of $\Omega$,  establishing variable selection consistency at the aggregated level.}
\textcolor{black}{
Furthermore, under appropriate assumptions on the sample size and data structure, our method guarantees that the classification results based on the sample are consistent with those from the underlying population distribution. Specifically, for each group $g$, define
$
\kappa_g := \vertiii{\Sigma_{g,S_g^CS_g}\Sigma_{g,S_gS_g}^{-1}}_{\infty}, \Lambda_g := \Theta_{S_g}^\top \Sigma_{g,S_gS_g}\Theta_{S_g}\in\mathbb{R}^{(G-1)\times(G-1)}
$
and let $\lambda_{\min}(\Lambda_g)$ denote the minimum eigenvalue of $\Lambda_g$. 
\begin{theorem}\label{Theorem:3}
Assume (A1) holds. Let the following conditions be satisfied:  
\textbf{(B1)} $n,p\to\infty$, $G=\mathcal{O}(1)$, $\tfrac{s^2\log(ps^2)}{n}\to0$, and 
$\mathcal{O}\!\big(\sqrt{\tfrac{s^2\log(ps^2)}{n}}\big)
<\mathcal{O}(\lambda)
<\mathcal{O}\!\big(\min_{g}\Theta_{g,\min}\big)$;  
\textbf{(B2)} $\inf_g\lambda_{\min}(\Lambda_g)>0$.  
If $\lambda\to0$ and $\kappa_g<1$ for each $g=1,\ldots,G$, then
\[
P\!\left(\hat{h}_{\widehat{\Omega}}(X)=h_\Theta(X)\right)\to1,
\]
where 
$
h_{\Theta}(X_{\text{new}})
=\arg\min_{1\le g\le G}
\{
(X_{\text{new}}-\mu_g)^{\top}\Theta
\left(\Theta^{\top}\Sigma_g\Theta\right)^{-1}
\Theta^{\top}(X_{\text{new}}-\mu_g)
+\log\!\left|\Theta^{\top}\Sigma_g\Theta\right|
-2\log\pi_g
\}
$
and 
$
\hat{h}_{\widehat{\Omega}}(X_{\text{new}})
=\arg\min_{1\le g\le G}
\{
(X_{\text{new}}-\bar{X}_g)^{\top}\widehat{\Omega}
(\widehat{\Omega}^{\top}\widehat{\Sigma}_g\widehat{\Omega})^{-1}
\widehat{\Omega}^{\top}(X_{\text{new}}-\bar{X}_g)
+\log\!\left|\widehat{\Omega}^{\top}\widehat{\Sigma}_g\widehat{\Omega}\right|
-2\log(n_g/n)
\}.
$
\end{theorem}
\begin{remark}\label{remark2}
Under the same assumptions, the estimator also has blockwise consistency on the active set $S=\cup_g S_g$, i.e.,
$
\vertiii{(\widehat{\Omega}-\Theta)_{S}}_{\infty,2}\to0,
$
which follows from the KKT conditions and the high-probability bounds in the proof of Theorem~\ref{Theorem:3}. 
Since the classifier depends on the basis only through projection basis, $\mu_g$ and $\Sigma_g$ (for $g=1,...,G$), this guarantees that the quadratic and log-determinant terms converge uniformly, leading to the classification consistency stated above.
\end{remark}}

\textcolor{black}{This theorem shows that when the sample size satisfies $n \gtrsim s^2 \log(ps^2)$, our method achieves both \emph{variable selection consistency} and \emph{classification consistency}; that is, the classifier based on the sample estimates converges to the population-level classifier. Beyond the Gaussian and full-rank setting, we also consider two extensions. First, we analyze the case in which the sample-based projection basis $\widehat{\Omega}$ is not of full rank and explicitly specify the corresponding classifier. Let $A_g := \widehat{\Omega}^\top \widehat{\Sigma}_g \widehat{\Omega}$. If $A_g$ is invertible, our method uses $A_g^{-1}$ and $\log|A_g|$ as in eq~\eqref{CLFR}. If $A_g$ is singular, we instead use the Moore--Penrose pseudoinverse $A_g^+$ and the pseudodeterminant $\operatorname{pdet}(A_g)$, defined as the product of its positive eigenvalues. Second, we study conditions under which our theoretical results extend to sub-Gaussian populations and show that, under suitable assumptions, the proposed method retains the same consistency guarantees. Detailed proofs of Theorems~\ref{Theorem:1}–\ref{Theorem:3}, together with the discussion of these two extensions, are provided in the Appendix Sections~\ref{proofs}–\ref{discussions}.}


\section{Simulation Studies}\label{sec4}
\subsection{Simulation Settings}
We evaluate the performance of MGQDA regarding both classification accuracy and variable selection. Prediction accuracy is measured by the misclassification rate, and variable selection is assessed using the true positive rate (TPR) and false positive rate (FPR). We compare MGQDA with the following methods:
shrinkage-mean-based diagonal QDA (SMDQDA, \citealp{tong2012improved}), high-dimensional regularized discriminant analysis (HDRDA, \citealp{ramey2016high}), multi-group sparse discriminant analysis (MGSDA, \citealp{gaynanova2016simultaneous}), discriminant analysis via projection (DAP, \citealp{gaynanova2019sparse}), sparse linear discriminant analysis (SLDA, \citealp{clemmensen2011sparse}), and penalized LDA (pLDA, \citealp{2011Penalized}). Since DAP is a two-group method, we extend it to multi-group problems by pairwise voting and denote it by ``GDAP''. Classical QDA based on the generalized inverse of the covariance matrix was initially considered, but it was excluded from the comparisons due to numerical instability and inflated error rates.


 In all simulations, we set the group-specific sample size to $n_g = 100$ for each $g$, consider $p \in \{200,500\}$ predictors, and examine $G \in \{3,5\}$ groups. Data are generated from multivariate normal distributions $(X \mid Y=g)\sim \mathcal{N}(\mu_g,\Sigma_g)$. To cover a range of covariance structures, we consider four families:
(i) \emph{block-equicorrelation}, with block size $b$ and parameter $\rho\in[0,1]$, where $\Sigma_g = \operatorname{diag}\{\rho I_b + (1-\rho)\iota_b\iota_b^{\top},\, I_{p-b}\}$ and $\iota_b$ is the $b$-vector of ones;  
(ii) \emph{block-autocorrelation}, with block size $b$ and parameter $\rho\in[0,1]$, where $\Sigma_g = \operatorname{diag}\{\widetilde{\Sigma}_b,\, I_{p-b}\}$ and $(\widetilde{\Sigma}_b)_{ij} = \rho^{|i-j|}$;  
(iii) \emph{spiked covariance}, with normalized vectors $q_1,q_2\in\mathbb{R}^p$ and scalars $a_1,a_2\in\mathbb{R}$, where $\Sigma_g = a_1 q_1q_1^{\top} + a_2 q_2q_2^{\top} + I_p$; unless stated otherwise, we set $q_1\propto(1,\ldots,b,0,\ldots,0)$ and $q_2\propto(b,\ldots,1,0,\ldots,0)$;  
(iv) \emph{block model}, where we let $\Lambda = \operatorname{diag}(\lambda_1,\ldots,\lambda_b)$ with $\lambda_i\sim\operatorname{Unif}[1,2]$ and $U$ a $b\times b$ matrix with i.i.d.\ standard normal entries, and set $\Sigma_g = \operatorname{diag}\{U^{\top}\Lambda U,\, I_{p-b}\}$.
We consider eight data-generating models. Models~1–5 use $G=3$ groups and Models~6–8 use $G=5$ groups. For each model, we examine $p=200$ and $p=500$. We set $b=30$ for Models~1–4 and $b=50$ for Models~5–8. The mean and covariance structures are:

\begin{itemize}
    \item[\textbf{Model 1:}] 
    $\mu_1 = \mathbf{0}_p$, $\Sigma_1$ is block-equicorrelation with $\rho = 0.8$; 
    $\mu_2 = (\mathbf{1}_{10}, \mathbf{-1}_{10}, \mathbf{0}_{p-20})$, $\Sigma_2$ is block-autocorrelation with $\rho = 0.8$; 
    $\mu_3 = (\mathbf{-1}_{10}, \mathbf{1}_{10}, \mathbf{0}_{p-20})$, $\Sigma_3$ is spiked with $(a_1,a_2)=(100,10)$.

    \item[\textbf{Model 2:}] 
    $\mu_1=\mathbf{0}_p$; 
    $\mu_2=(\mathbf{0.5}_{10},\mathbf{-0.5}_{10}, \mathbf{0}_{p-20})$; 
    $\mu_3=(\mathbf{-0.5}_{10},\mathbf{0.5}_{10}, \mathbf{0}_{p-20})$; 
    the covariance matrices $(\Sigma_1,\Sigma_2,\Sigma_3)$ are the same as in Model~1.

    \item[\textbf{Model 3:}]
    $\mu_1 = \mathbf{0}_p$, $\Sigma_1$ is block-equicorrelation with $\rho = 0.3$; 
    $\mu_2 = (\mathbf{1}_{10}, \mathbf{-1}_{10}, \mathbf{0}_{p-20})$, $\Sigma_2$ is block-autocorrelation with $\rho = 0.7$; 
    $\mu_3 = (\mathbf{-1}_{10}, \mathbf{1}_{10}, \mathbf{0}_{p-20})$, $\Sigma_3$ is from the block model with $U,\Lambda$ generated as above.

    \item[\textbf{Model 4:}]
    $\mu_1 = (\mathbf{1}_{10}, \mathbf{0}_{p-10})$, $\Sigma_1$ is block-equicorrelation with $\rho = 0.3$; 
    $\mu_2 = (\mathbf{0}_{10}, \mathbf{1}_{10}, \mathbf{0}_{p-20})$, $\Sigma_2$ is block-autocorrelation with $\rho = 0.7$; 
    $\mu_3 = (\mathbf{0}_{20}, \mathbf{1}_{10}, \mathbf{0}_{p-30})$, $\Sigma_3$ is spiked with $(a_1,a_2)=(30,5)$, and here we set $q_1 \propto (1, \dots, \sqrt{b}, 0, \dots, 0)$ and $q_2 \propto (\sqrt{b}, \dots, 1, 0, \dots, 0)$.

    \item[\textbf{Model 5:}]
    Let $\rho_0 = 0.8$. Then 
    $\mu_1 = (\mathbf{1}_{10}, \mathbf{0}_{p-10})$, $\Sigma_1$ is block-autocorrelation with $\rho = \rho_0$; 
    $\mu_2 = (\mathbf{0}_{10}, \mathbf{1}_{10}, \mathbf{0}_{p-20})$, $\Sigma_2$ is block-autocorrelation with $\rho = 0.7\rho_0$; 
    $\mu_3 = (\mathbf{0}_{20}, \mathbf{1}_{10}, \mathbf{0}_{p-30})$, $\Sigma_3$ is block-autocorrelation with $\rho = 0.3\rho_0$.

    \item[\textbf{Model 6:}]
    $\mu_1 = (\mathbf{1}_{10}, \mathbf{0}_{p-10})$, $\Sigma_1$ is block-equicorrelation with $\rho = 0.5$; 
    $\mu_2 = (\mathbf{0}_{10}, \mathbf{1}_{10}, \mathbf{0}_{p-20})$, $\Sigma_2$ is block-autocorrelation with $\rho = 0.5$; 
    $\mu_3 = (\mathbf{0}_{20}, \mathbf{1}_{10}, \mathbf{0}_{p-30})$, $\Sigma_3$ is spiked with $(a_1,a_2)=(100,10)$; 
    $\mu_4 = (\mathbf{0}_{30}, \mathbf{1}_{10}, \mathbf{0}_{p-40})$, $\Sigma_4$ is spiked with $(a_1,a_2)=(10,100)$; 
    $\mu_5 = (\mathbf{0}_{40}, \mathbf{1}_{10}, \mathbf{0}_{p-50})$, $\Sigma_5$ is from the block model with $U,\Lambda$ generated as above.

    \item[\textbf{Model 7:}]
    The covariance matrices are the same as in Model~6. The mean vectors are scaled down: $\mu_g^{(7)} = 0.5\,\mu_g^{(6)}$ for $g=1,\ldots,5$, where $\mu_g^{(6)}$ denotes the mean in Model~6.

    \item[\textbf{Model 8:}]
    The covariance matrices are again the same as in Model~6. The mean vectors are  
    $\mu_1=\mathbf{0}_p$; 
    $\mu_2=(\mathbf{1}_{20}, \mathbf{0}_{p-20})$; 
    $\mu_3=(\mathbf{0.2}_{10}, \mathbf{-0.2}_{10}, \mathbf{0}_{p-20})$; 
    $\mu_4=(\mathbf{0}_{20}, \mathbf{5}_{2}, \mathbf{0}_{p-22})$; 
    $\mu_5=(\mathbf{0}_{25}, \mathbf{-10}, \mathbf{10}, \mathbf{0}_{p-27})$.
\end{itemize}

For each method, the empirical classification error rate on an independent test set is used to measure prediction accuracy. 
\textcolor{black}{Details on the calculation of variable-selection metrics (TPR/FPR) and group-specific active sets are provided in Section~\ref{Simres} and Appendix Section \ref{supp5}.}

\subsection{Results}\label{Simres}

\begin{figure}[htbp]
    \centering
    \includegraphics[width=0.9\textwidth]{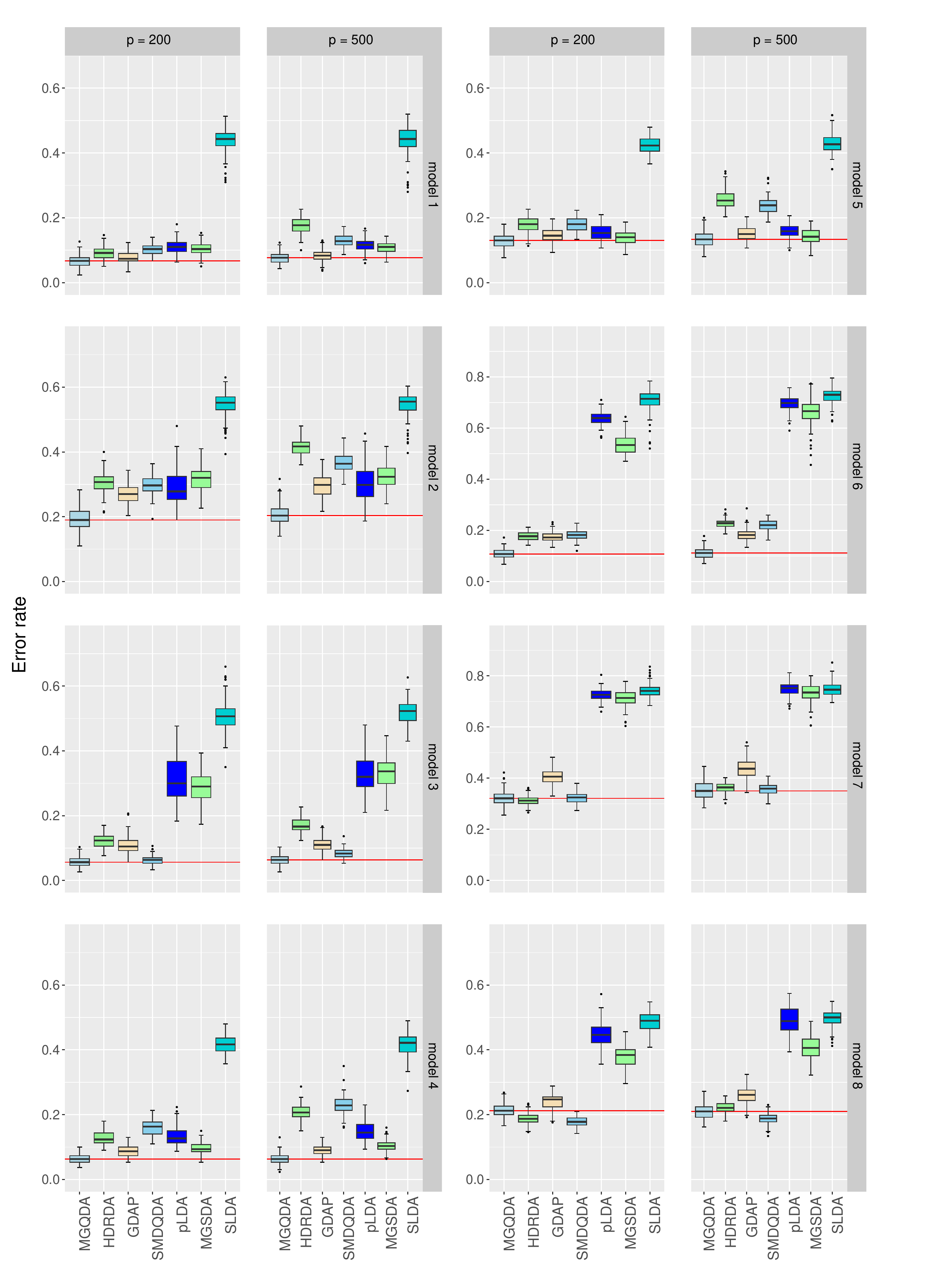} 
    \caption{\footnotesize 
    Misclassification error rates for MGQDA and competing methods across all simulation settings, based on 100 replications. The horizontal line in each panel marks the median error of MGQDA.}
    \label{errorrate}
\end{figure}

\textcolor{black}{
Overall, MGQDA performs strongly across nearly all simulation settings, typically achieving the lowest or near-lowest misclassification rates among the methods
 considered (see Figure~\ref{errorrate}).  Comparing Models~1 and~2 (and similarly Models~6 and~7) shows that MGQDA, HDRDA, GDAP, and SMDQDA remain competitive even when the mean differences are relatively weak, indicating that these procedures effectively exploit differences in covariance structure. In contrast, when the within-group covariance matrices become more homogeneous across classes (e.g., in Model~5), the advantage of fully quadratic rules naturally diminishes and the relative performance of all methods becomes more similar.}

\begin{figure}[htbp]
    \centering
    \includegraphics[width=\textwidth]{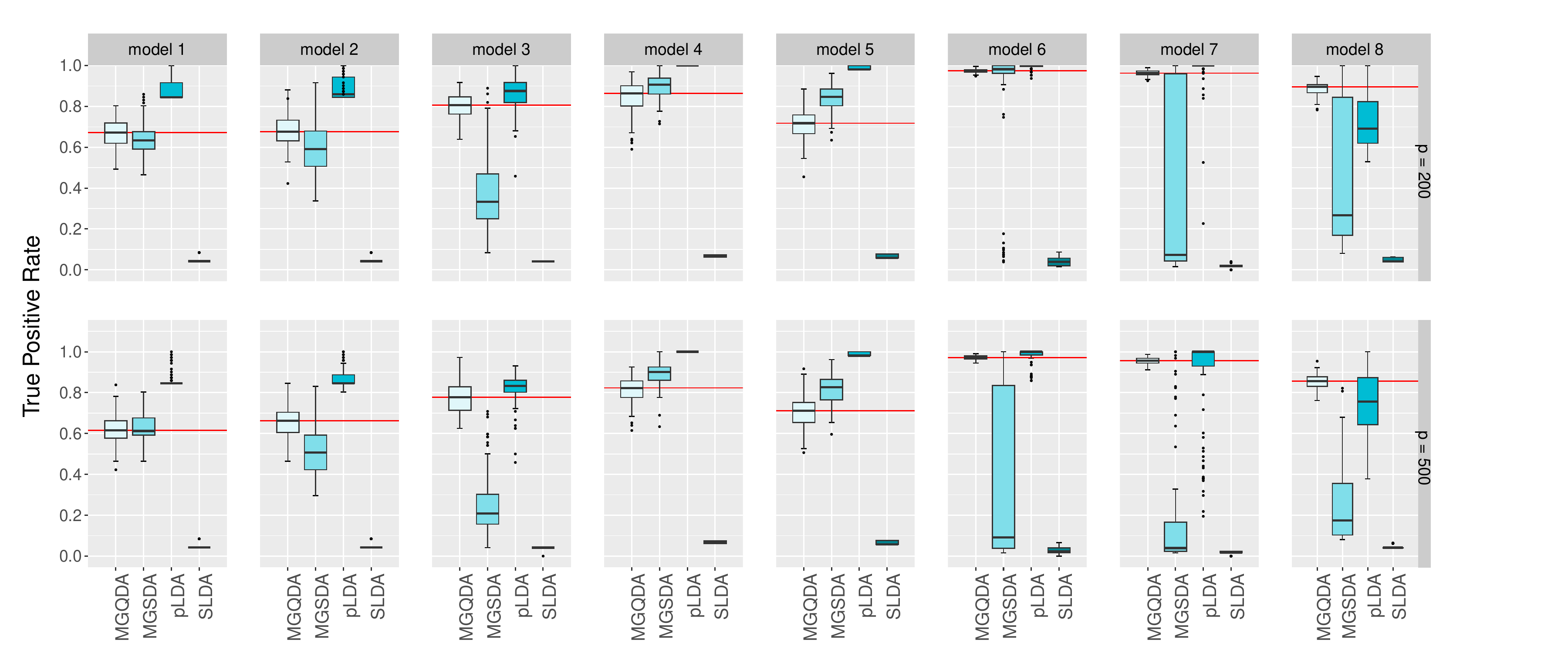} 
    \caption{\footnotesize TPR for MGQDA, MGSDA, pLDA, and SLDA across all simulation settings, based on 100 replications. The red horizontal line in each panel marks the median TPR of MGQDA.}
    \label{TPRplot}
    \includegraphics[width=\textwidth]{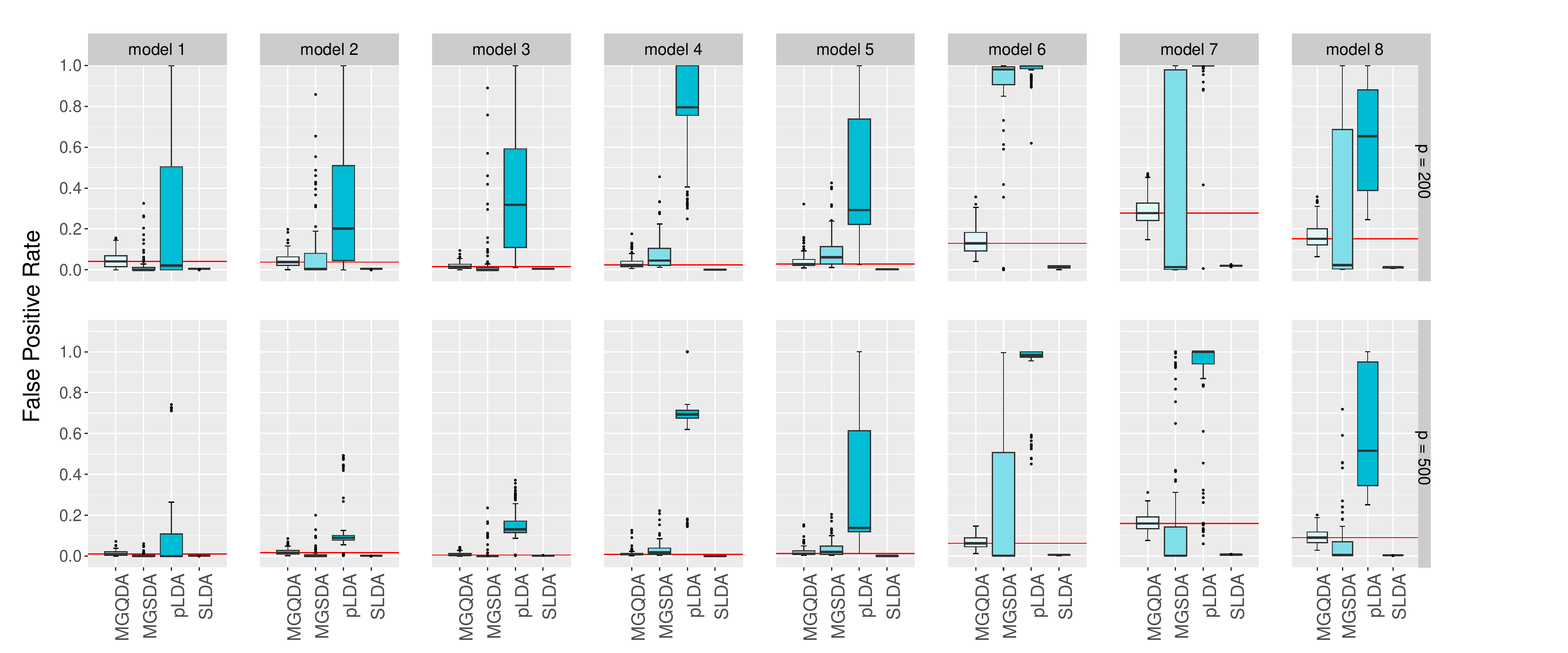} 
    \caption{\footnotesize FPR for MGQDA, MGSDA, pLDA, and SLDA across all simulation settings, based on 100 replications. The red horizontal line in each panel marks the median FPR of MGQDA.}
    \label{FPRplot}
\end{figure}

{\color{black}
Figures~\ref{TPRplot} and~\ref{FPRplot} summarize the overall variable-selection 
performance of the sparse methods (MGQDA, MGSDA, pLDA, and SLDA) by reporting 
the average TPR and FPR across 
100 replications.  
MGQDA achieves the most favorable balance between sensitivity and specificity in 
nearly all simulation settings. Its TPR remains high and stable as the dimension 
increases from $p=200$ to $p=500$, while its FPR stays uniformly low, indicating 
accurate recovery of active variables without excessive false selections.
}

{\color{black}
By contrast, MGSDA exhibits moderate TPR with noticeably larger variability and 
higher FPR, reflecting less stable support estimation.  
pLDA often attains relatively high TPR but suffers from substantially inflated 
FPR, suggesting that although many true signals are detected, many noise variables 
are also included.  
SLDA tends to produce overly sparse models, resulting in extremely low TPR and 
limited discriminative utility.  
Together these comparisons show that MGQDA provides the most stable and reliable 
support recovery under heterogeneous covariance structures.
}

{\color{black}
To further examine variable selection at the group level, detailed group-specific 
TPR and FPR results and explicit construction of both true and estimated active sets 
are provided in Appendix~\ref{supp5}.  
Using Setting~3 as an illustrative example, Appendix~\ref{supp5} demonstrates 
how the support sets $S_g$ and $\widehat{S}_g$ are formed for each group and 
shows that MGQDA maintains robust performance across heterogeneous groups, 
consistent with the aggregated results reported in Figures~\ref{TPRplot}–\ref{FPRplot}.
}

{\color{black}
Combined with the misclassification results, these findings show that MGQDA
consistently achieves the most favorable combination of predictive accuracy and
variable-selection reliability across all models, dimensions, and covariance
structures.  
This empirical evidence supports the theoretical guarantees in
Section~\ref{sec3}.
}

\spacingset{1}
\spacingset{1.75}

\section{Real Data Application}\label{sec5}

\textcolor{black}{
We evaluate the performance of our method using gene expression data from the DepMap Public 23Q2 release \citep{depmap23q2,arafeh2025present}, which provides large-scale genomic and functional profiles of cancer cell lines and is widely used to study genetic dependencies and vulnerabilities in cancer \citep{killian2021exploiting,bazaga2020genome,chiu2021predicting}. The dataset contains RNA-seq–based expression measurements for 1,864 cancer cell lines, reported as $\log_2(\text{TPM}+1)$-transformed values. In total, expression levels are available for 19,193 protein-coding genes, offering a comprehensive and standardized resource for evaluating the proposed classifier.
}

Due to the high dimensionality of the data, we first applied pairwise $t$-tests to screen for influential genes and retained $p = 800$ genes for downstream analysis. Similar screening strategies have been used in \citet{gaynanova2016simultaneous,2011Penalized,gaynanova2019sparse}. We then benchmarked MGQDA against other multi-group discriminant analysis methods, including SMDQDA, HDRDA, MGSDA, and pLDA. We excluded SLDA because it performed substantially worse in our simulation studies, and we did not include DAP here because it is inherently a two-class method (our simulations used a pairwise voting extension, ``GDAP,'' only for completeness).


For the main analysis, we focused on the five diseases with the largest numbers of primary lesion samples: ``Non-Small Cell Lung Cancer,'' ``Non-Hodgkin Lymphoma,'' ``Head and Neck Squamous Cell Carcinoma,'' ``Diffuse Glioma,'' and ``Colorectal Adenocarcinoma.'' Each disease contributed between 40 and 77 samples. We fixed $p = 800$ across all methods to facilitate direct comparison. To assess robustness, we also carried out three additional experiments using different disease subsets and different values of $p$; detailed results are reported in the Appendix~\ref{supp4}.


Figure~\ref{realdata} summarizes performance across 100 replications. In each replication, 90\% of the samples were randomly assigned to a training set and the remaining 10\% to a test set. Classifiers were trained on the genetic data and disease labels in the training set and evaluated on the held-out test set. The proposed MGQDA method achieves lower classification error rates than the competing approaches (Figure~\ref{realdata}, left). HDRDA performs comparably in terms of misclassification but does not perform variable selection.


\textcolor{black}{Regarding the variable selection results, MGQDA consistently selects a much sparser model than MGSDA (Figure~\ref{realdata}, middle), with a substantial portion of features selected only for certain groups, highlighting the ability of our method to identify structural differences.} HDRDA, SMDQDA, and pLDA use all $p$ variables and are therefore omitted from the variable-selection plots. Among the genes most frequently selected by MGQDA, several have strong support in the existing literature. In the ``Non-Small Cell Lung Cancer'' group, genes \textcolor{black}{DDR2}, PLEK2, and CDH13 are among the top selected, consistent with their reported roles in \citet{sato1998h}, \citet{shen2019plek2}, and \citet{hammerman2011mutations}, respectively. In the ``Diffuse Glioma'' group, RAB34 is frequently selected, in line with its documented relevance in glioma biology \citep{wang2015rab34}.


\spacingset{1}
\begin{figure}[htbp]
    \centering
    \includegraphics[width=0.95\textwidth]{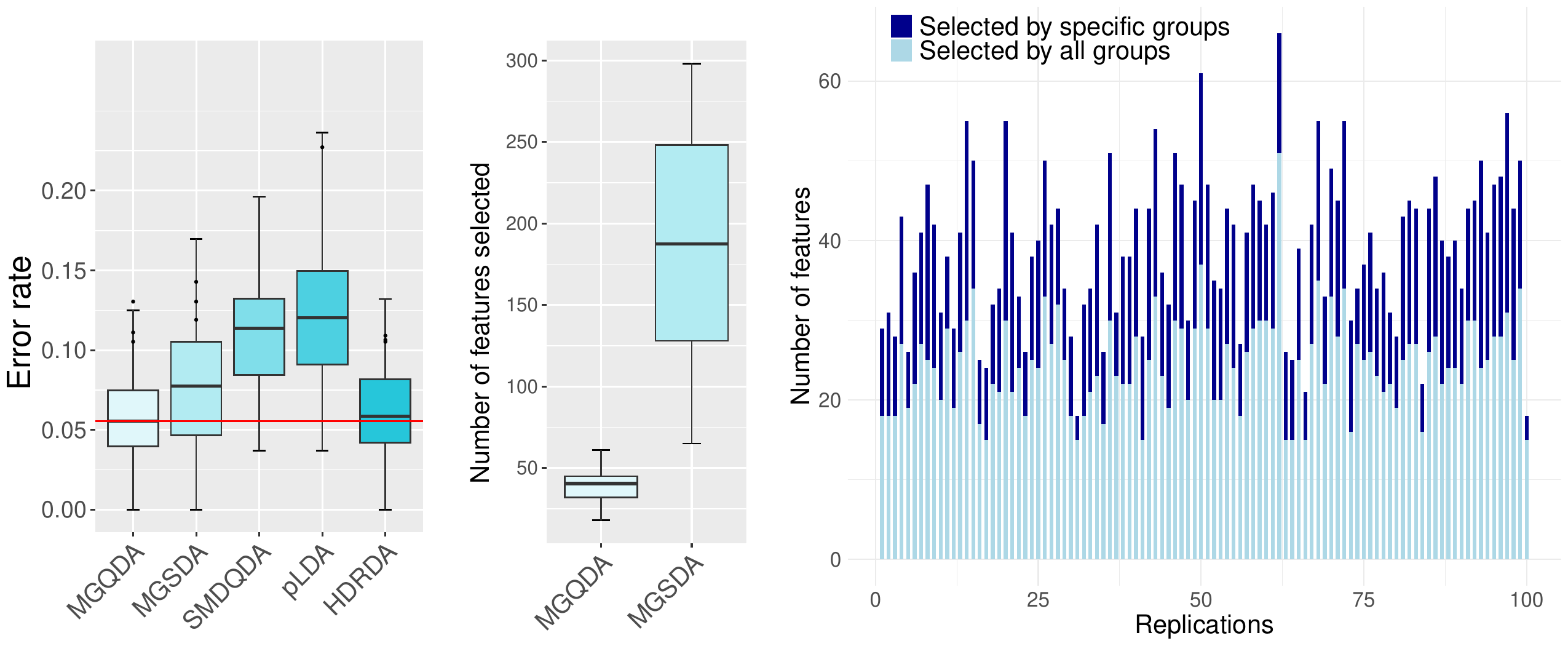} 
    \caption{\footnotesize Classification performance and variable selection on the DepMap gene expression dataset for MGQDA and competing methods. Left: classification error rates across 100 replications (red horizontal line: MGQDA median error). Middle: number of selected variables across replications for sparse methods. Right: MGQDA-selected variable counts by experiment; the $x$-axis indexes experiments and the $y$-axis gives the number of variables selected in the projection basis.
    }\label{realdata}
\end{figure}
\spacingset{1.75}

Therefore, MGQDA achieves lower classification error rates while selecting sparse sets of variables that capture both shared and group-specific structure. This leads to interpretable classification rules, and the biological relevance of the top-selected genes is supported by existing cancer genomics studies.

\section{Discussion}
\label{sec6}

We propose a new approach to multi-group discriminant analysis that accommodates distinct covariance structures across groups while retaining effective variable selection and practical computational cost. Extensive simulation studies (Section~\ref{sec4}) and the real-data analysis (Section~\ref{sec5}) show that MGQDA can substantially improve classification accuracy in settings with heterogeneous covariance matrices. In addition to improved predictive performance, the method remains computationally efficient, with complexity that scales linearly in $p$ for fixed $G$ and $n$, making it more practical than procedures that rely on direct high-dimensional covariance estimation. A key advantage of our framework is its ability to perform group-specific variable selection, as illustrated in the DepMap analysis, where MGQDA identifies both shared and disease-specific markers with clear interpretability. One natural extension is to incorporate shrinkage strategies, in the spirit of regularized discriminant analysis, to improve robustness when some groups have very small sample sizes. We leave these developments and broader applications of MGQDA in large-scale studies to future work.


{\color{black}

\section*{Supplementary Materials}

\begin{description}

\item[Appendix:] Notation and technical preliminaries; detailed proofs for all Propositions and Theorems (Theorem \ref{Theorem:1}--\ref{Theorem:3}) with auxiliary lemmas; additional explanations of experimental settings; and additional results for the data analysis.

\item[R code:] The R code for implementing the MGQDA proposed in the paper.

\item[Data Availability Statement:] The DepMap Public 23Q2 data analyzed in this study (figshare DOI: 10.6084/m9.figshare.22765112.v2) are publicly available via the DepMap portal (https://depmap.org/portal). The dataset contains cell‑line–level measurements only and does not include personally identifiable information.
\end{description}

\bibliographystyle{chicago}
\bibliography{ref}
\end{document}